# Role of Edge Device and Cloud Machine Learning in Point-of-Care Solutions Using Imaging Diagnostics for Population Screening

Dr Amit Kharat[1], Dr Vinay Duddalwar[2], Dr Krishna Saoji[3], Ashrika Gaikwad[1], Viraj Kulkarni[1], Gunjan Naik[1], Rohit Lokwani[1], Swaraj Kasliwal[1], Sudeep Kondal[1], Tanveer Gupte[1], Aniruddha Pant[1]

[1] DeepTek Inc
[2] University of Southern California, USA
[3] DY Patil Hospital, Pune

*Abstract*- Edge devices are revolutionizing diagnostics. Edge devices can reside within or adjacent to imaging tools such as digital X-ray, CT, MRI, or ultrasound equipment. These devices are either CPUs or GPUs with advanced processing deep and machine learning (artificial intelligence) algorithms that assist in classification and triage solutions to flag studies as either normal or abnormal, TB or healthy (in case of TB screening), suspected COVID-19/other pneumonia or unremarkable (in hospital or hotspot settings). These can be deployed as screening point-of-care (PoC) solutions; this is particularly true for digital and portable X-ray devices. Edge device learning can also be used for mammography and CT studies where it can identify microcalcification and stroke, respectively. These solutions can be considered the first line of pre-screening before the imaging specialist actually reviews scans and makes a final diagnosis. The key advantage of these tools is that they are instant, can be deployed remotely where experts are not available to perform pre-screening before the experts actually review, and are not limited by internet bandwidth as the nano learning data centers are placed next to the device.

*Index Terms*- Edge device, artificial intelligence, edge learning, population screening, deep learning, machine learning, cloud learning, teleradiology, point-of-care solutions.

I. INTRODUCTION

Edge device learning is creating a wave of transformation in the healthcare sector. It promotes the deployment of deep and machine learning tools at the edge of a network [1]. An edge device can classify and localize disease conditions from radiology images like digital X-ray, ultrasound, mammography, CT, and MRI. By definition, an edge device is a system that allows computing to happen closer to the location where it is needed, resulting in an instant response by saving essential bandwidth [2]. According to Gartner, edge computing involves carrying out content aggregation, information processing, and delivery close to the information sources to reduce latency and keep traffic local and distributed [3].

When edge computing is applied to medical imaging devices, it reduces the time needed for diagnosis by instant triage and thereby optimizes the diagnostic workflow. It proposes a system that allows a set of conditions to be pre-screened and flagged by processing the medical imaging dataset on the spot without any need to push the images to a cloud network to get analyzed by a cloud-based algorithm before it is reviewed by a human expert. Edge learning devices are also called nano data centers. They can work in harsh weather conditions with minimal to no requirement of air conditioning. Their key advantages include non-dependency on internet bandwidth and zero latency issues as they are located adjacent to the peripherally deployed medical device. However, nano data centers suffer from a low computation capacity and the number of algorithms that can be deployed on the edge device is limited.

II. THE HUMAN COGNITIVE PROCESS OF ANALYSIS OF MEDICAL IMAGING FINDING

Lesions seen on digital X-ray, CT, MRI, ultrasound, or mammography images follow certain patterns. These patterns are considered pathognomonic of a given condition. When a particular pattern of distribution or appearance is present and manifestations on an image are apparent, a specific list of diagnoses or differentials is given by an imaging expert. The list of differential diagnoses or a definitive single diagnosis depends on certain imaging pathognomonic features and inputs from biochemical and microbiological tests apart from clinical inputs. The cognitive process of gathering information for a medical image is complex and experts access the information by reference to prior visual cues, memory, in-depth understanding of academic literature, and subject matter expertise collated through a complex process of learning. The process has a strong association with the experiential knowledge that imaging experts have gathered over the years after sifting through many images and understanding the specific clinical conditions that can lead to a particular diagnosis.

Analogously, an algorithm can be devised to perform pattern recognition using machine learning and artificial intelligence. Using supervised, unsupervised and semi-supervised learning models, an attempt is made to replicate human cognition. This is called narrow artificial intelligence (AI), where machines can be programmed to perform specific tasks, one task at a time. For example, take the diagnosis of COVID-19 or community-acquired pneumonia from a digital chest X-ray or CT image (Figure 1 and Figure 2). After sufficient training, a model can potentially detect COVID-19 or other types of community-acquired pneumonia patterns from either CT or X-ray images. However, detecting a pattern does not necessarily mean the disease condition exists, and a similar pattern can also be seen in other community-acquired types of pneumonia. Nonetheless, the algorithm can flag and triage studies and present them to the expert for review and analysis and redirecting these studies to a faster workflow which allows for study prioritization and scaling up or down on the dynamic multimodality worklist accessed by imaging experts. However the same algorithm may not be able to diagnose or detect TB as it has not been trained for this new task, detecting TB from a digital chest X-ray becomes another narrow AI solution, where deep learning algorithms can pre-screen and flag from a digital chest X-ray (Figure 3).

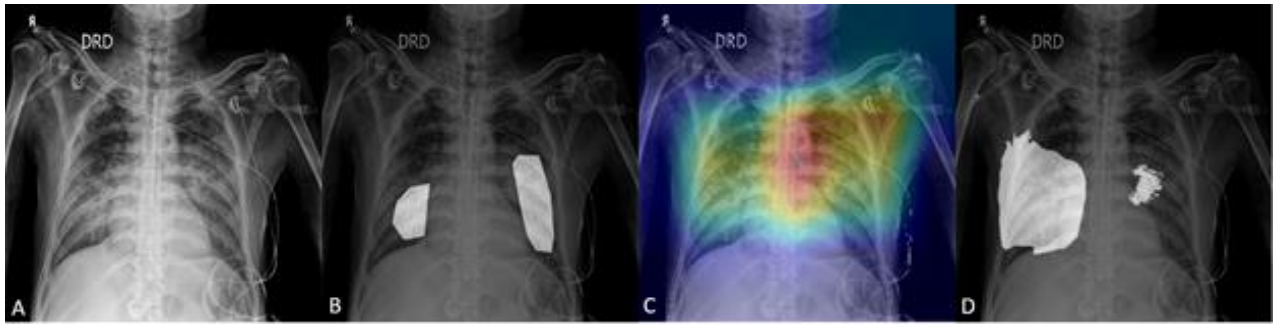

Figure 1: (A) Original Chest X-ray image; (B) radiologist annotations; (C) predicted heatmap; and (D) predicted mask for COVID-19 detection.

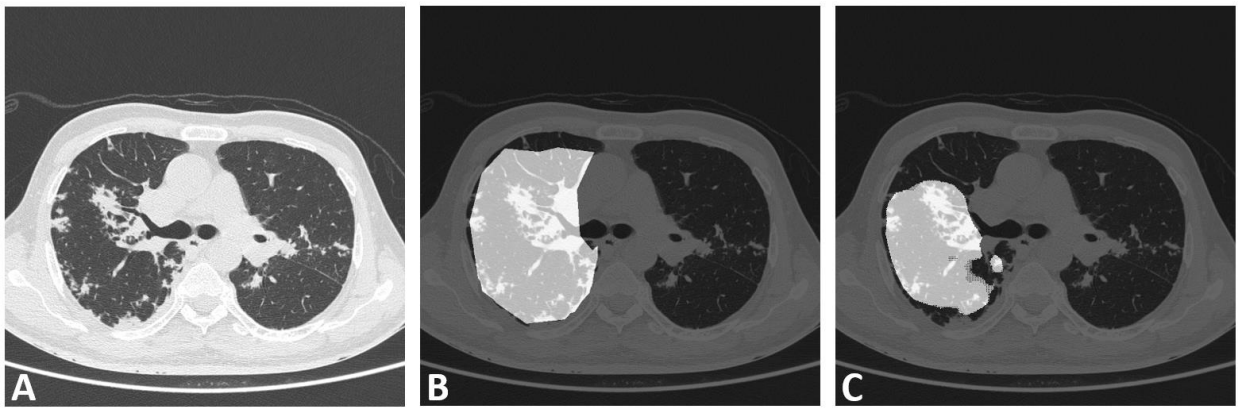

Figure 2: (A) Original CT image; (B) radiologist's annotated mask; and (C) mask predicted by the model for lung inflammatory process.

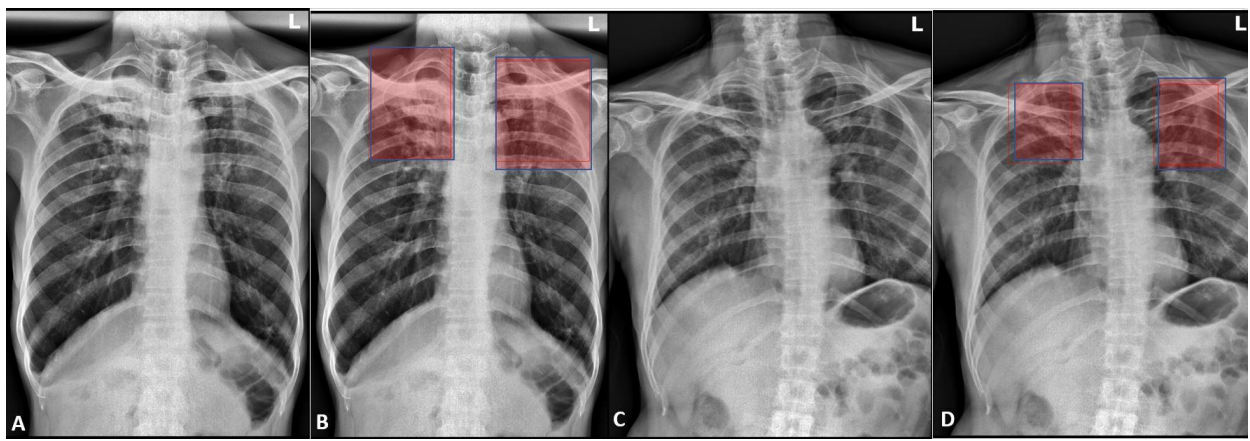

Figure 3: (A) Original chest X-ray image for patient one; (B) predicted mask image for TB for patient one; (C) Original X-ray image for patient two; and (D) predicted mask image for TB for patient two.

III. EDGE SOLUTIONS

Edge solutions can be used for population screening of predefined disease conditions. Few such examples are listed below (Table 1):

*A. Edge devices can screen the following conditions using chest X-rays:*
1. Tuberculosis (TB): active or old (latent TB)
2. Pneumonia related to COVID-19, Severe Acute Respiratory Illness Syndrome (SARS), and H1N1 influenza or swine flu
3. Nodules for lung cancer
4. Pneumothorax.

*B. Edge devices can screen the following from mammography images:*
1. Microcalcifications
2. Mass.

*C. Edge solutions, when deployed on CT, can be used for the following conditions:*
1. CT Chest:
    a) Pneumonia detection (COVID-19, community acquired pneumonias etc.)
    b) Tuberculosis detection
    c) Interstitial lung disease detection (ILD)
    d) Lung mass and lung nodules.
2. CT Brain:
    a) Stroke detection
    b) Excluding bleed and detecting acute infarcts.

*D. For MRI, the following solutions can benefit from edge tools:*
1. MRI Brain
    a) Stroke detection (acute infarct or bleed detection).

| Edge Solutions | **X-ray Chest** | **Disease** | **Subtypes** |
|---|---|---|---|
| | | TB | Active/Old/Others |
| | | Pneumonia | COVID-19/SARS/H1N1 |
| | | Cancer Lungs | Nodule |
| | | Emergency | Pneumothorax |
| | Mammography | | |
| | | Microcalcifications | |
| | | Mass | |
| | CT Chest | | |
| | | Pneumonia | COVID-19 |
| | | Other Infections | TB |
| | | Inflammatory | Interstitial Lung Disease |
| | CT Brain | | |
| | | Stroke | Bleed Detection |
| | MRI Brain | | |
| | | Stroke | Infarct Detection |

Table 1: Examples of AI solutions that can be potentially deployed on edge devices.

## IV. IMAGING DEVICES THAT QUALIFY AS EDGE SOLUTION

As a screening tool, the device on which the algorithm is deployed also needs to have certain characteristics. The screening device enabled with deep learning tools should ideally be portable and ready for quick deployment in the suburbs and rural areas. Mobile Diagnostic Units (MDUs), which hosts X-rays or mammography machines, can meet these requirements. Additional key features for edge devices to have an impact on diagnostics include operability of the diagnostic devices on a single-phase connection or on battery or reserve power. Ability to deliver good quality images in digital DICOM format consistently. DICOM images should be seamlessly transferable across workstations and other DICOM readers or cloud platforms for further analysis and quality checks. Furthermore, the personnel operating the devices should be able to manage the equipment with adequate training and follow protocols to ensure minimal exposure to radiation and compliance with specific operating standards.

Many digital imaging devices may not fit the above criteria but can still be edge learning compatible. Any device that can generate DICOM data can potentially be used to assist in augmenting healthcare solutions for experts. In such an environment, the edge solutions process data in real-time with no dependency on network breakdown or congestion [4]. This is the likely last-mile approach to make healthcare diagnostics more accessible.

## V. TB SCREENING AND EDGE LEARNING SOLUTIONS AS PART OF POINT-OF-CARE SOLUTIONS

In population health, TB figures as the key disease condition where digital chest X-rays play a pivotal role in disease assessment and management. The Indian government has set 2025 as the goal to eliminate TB, and the World Health Organization plans to do the same by 2030 [5][6].

The current model of operations without AI or edge devices works as follows.

X-ray diagnostic equipment is used either in a hospital environment or in MDUs operating in the outskirts of a city. Images obtained from these are either pushed by encrypted protocol to be read by an expert remotely or physically reported by taking hard copy prints of images.

In a conventional system of digital reporting on a standard Picture Archiving and Communication System (PACS), a study cannot be prioritized unless the imaging expert either reads and reports all studies in the order in which they were obtained or manually searches for abnormal studies. This typical workflow does not allow imaging experts to prioritize the evaluation of studies with potentially urgent findings or abnormalities. Thus, workflow management becomes challenging and non-optimized and leads to reporting delays, stress, and burnouts amongst caregivers. Point-of-care solutions that allow testing and pre-screening near the patient mitigate these issues as they employ data-driven technologies to promote mobile healthcare. [7]

AI models help in triaging and prioritizing studies on the reading list and upscaling or downscaling the studies on the worklist for the radiologists and imaging experts. Additionally, the AI models can flag studies instantaneously so that patients can be held back for additional investigations. This reduces the number of patients lost to follow-up and the cost of follow-up. It also assists in controlling the spread of infection to other people in society. The overall burden of stress on imaging experts and other healthcare workers is also significantly reduced. This system can potentially perform COVID-19 screening in hospital settings or pneumothorax and pulmonary edema screening in ICU settings.

Porting these models on edge devices allows us to deploy narrow AI solutions remotely, and this is a gamechanger in population health screening. If the entire triage solution is available as an edge solution, it can be added in the workflow to raise alerts and enable smart notifications. Having such a solution in a TB screening program has significant value as instant triage ensures patients are not lost to follow-up.

## VI. LOST TO FOLLOW-UP AND COST BURDEN TO SOCIETY

Building a platform with triage solutions enables us to have an "end to end workflow solution", which ensures smooth documentation of the workflow, audit trails, and tracking of patients for follow-up visits, as smart notifications can be generated after a fixed period to remind patients of follow-up clinical examinations. Taking the TB example ahead, it may be noted that the provision of care gap between patients diagnosed with smear-positive TB and those who fail to start treatment is significant [8][9]. Using edge device-assisted triage and flag solutions, we can prevent pre-diagnostic loss to follow up and pre-diagnostic default of patients so that they can be flagged to undergo further testing by automated nucleic acid molecular diagnostics. Follow-up can also be done with patients on the phone to ensure that the patient does not default. The system can also potentially connect with existing government databases in respective countries for disease monitoring. Nikshay is one such platform in India that serves as the National Tuberculosis Surveillance System [10]. Connecting web-based radiology end-to-end workflow solutions backed by Artificial Intelligence with Nikshay or similar platforms in other nations can assist in augmented web-based patient surveillance tracking and monitoring in the specific geographic regions where TB is endemic.

## VII. INFECTIOUS DISEASES AND THE CLOUD MODEL

If a mature algorithm is translocated on the cloud, the results of X-ray screening will be available as and when the image gets uploaded to the cloud. This depends on the availability of the internet at that time. Typically, by using secure encrypted technology and a secure mobile network, it is possible to load thousands of images to the cloud for a near instantaneous triage by a trained algorithm.

Problems arise when the absence of internet connectivity spells a delay in the generation of triage and results arrive after the patient has already exited the X-ray imaging facility. For organizations hoping to curb and eliminate disease conditions like TB, a potential patient who has moved out of the facility means a lost opportunity to instantly triage and inability to quickly direct the patient for additional clinical, laboratory or molecular tests in the same setting.

A patient who is lost to follow-up takes a large toll on society in terms of additional costs to trace the patient and the possibility of the patient infecting others [11]. If disease elimination and disease eradication are the goals, a sharp focus is needed to monitor every positive study and mitigate risks. An edge device, acting as a virtual trained expert available in the form of an algorithm, can analyze and pre-screen the disease conditions without requiring internet and cloud solutions. This can be the "expedient last-mile approach" much needed in point-of-care diagnostics to eradicate conditions like TB.

Using a cloud-deployed algorithm, however, has distinct advantages. The AI algorithm residing in the cloud model uses the high computing power on-demand to run AI-based triage for many conditions apart from the disease in question. For example, a model deployed for TB triage can also help screen for cardiomegaly, foreign body, lung nodule/mass and assess quality parameters on X-ray images. The Cloud model also allows quick repurposing of a solution to detect TB to shift attention to other pathologies like community-acquired pneumonia (COVID-19 or other inflammatory lung conditions ) pattern assessment. This may not be possible on the nano data center on an edge device due to its limited computing capacity. Figure 4 lists the advantages and drawbacks of both solutions.

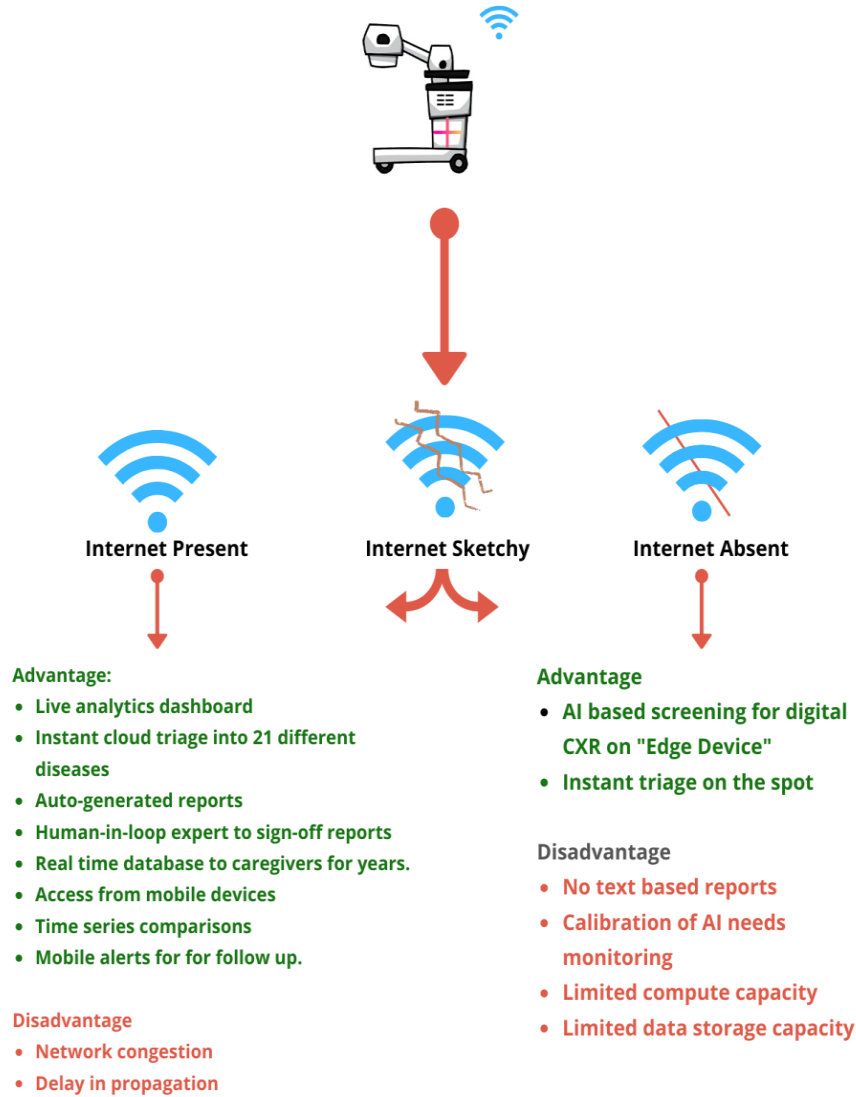

Figure 4: The difference between cloud and edge solutions.

An example of the MDU solution can be seen in a collaborative effort by the Government of Haryana and Medanta – The Medicity, a private hospital in Gurugram [12]. As part of the TB-Free Haryana campaign, a van fitted with digital X-ray equipment was deployed to a government health facility in the state of Haryana every week. Their results supported the use of MXUs (mobile X-ray units) in screening and active case finding for TB, owing to their operational and economic viability.

Similarly, the Greater Chennai Corporation's active screening for TB under the TB Free Chennai Project, using mobile diagnostic vans (MDUs) and the DxTB solution from Deeptek, is another example of deployment to pre-screen and triage for TB [13]. The AI triage results were reviewed by an imaging expert in the loop and further validated. The model assessed 75,000 X-ray images of the patients and classified each of them into one of two classes: (1) patient likely to have TB and (2) patient unlikely to have TB. All these images were later evaluated by an expert radiologist. When the model predictions were compared to the radiologist readings, the model demonstrated an AUROC of 0.97, 95% CI [0.94,0.99], and a sensitivity of 0.90, 95% CI [0.81,0.98], at a specificity of 0.95, 95% CI [0.93,0.96], (see Figure 5).

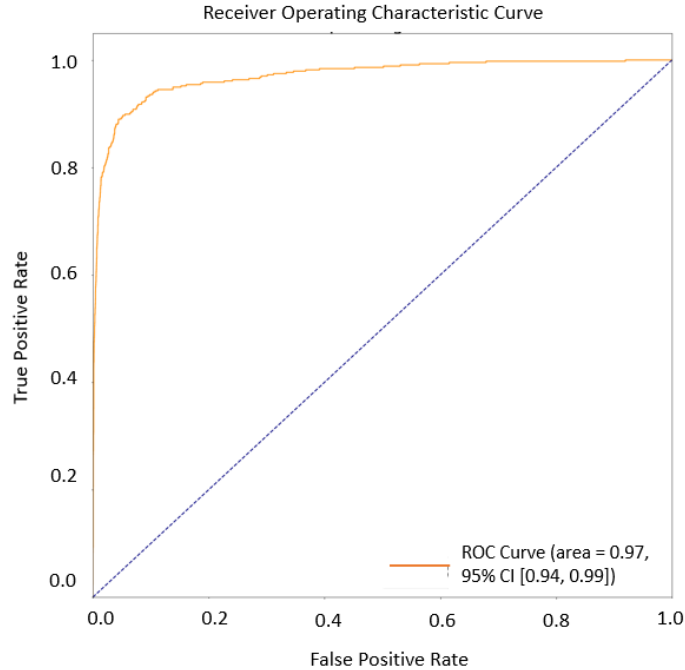

Figure 5: Performance metrics of the AI-based model for TB screening in the population screening setting.

Advantages of the cloud model: cloud technology allows batch processing of data. It works by engaging an imaging expert in the loop. Imaging experts/radiologists can review without having to worry about the software and hardware components of the system, as computation is not hosted locally, and they can use the regular workstations to review the images from remote location [14]. Correcting a drift in the deep learning algorithm also becomes easier. The cloud is flexible to allow review of other disease conditions and not just the disease under watch. For example, if cardiomegaly or lung mass is present in a patient having TB, it cannot be detected on a nano data center trained to identify and prescreen only TB. Cloud computing can also handle additional algorithms running simultaneously to process the X-ray for other pathologies simply because of higher computing capacity. The cloud model for AI-based triage also enables flagging of subtle variations in the disease conditions.

## VIII. DESCRIBING A TYPICAL EDGE DEVICE

A typical edge device can be as basic and cost-effective as the Raspberry Pi or a simple CPU that powers radiology workstations and desktops in its basic iteration, or it can be a high-end device as the NVIDIA A100 and AMD Radeon GPUs [15]. Placing a portable X-ray imaging unit equipped with a CPU/GPU in mobile diagnostic vans and in imaging clinics or hospitals is a game-changer. Equipment becomes smarter and healthcare more accessible. It increases operational excellence as well as empowers end-user (technologist or healthcare worker) engagement. Basic triage, pre-screening and pre-diagnosis can be performed by these devices before seeking specialist level intervention. These edge devices have the potential to act as point-of-care devices, capable of limiting epidemics and pandemics if deployed systematically.

## IX. HYBRID MODELS: EDGE + CLOUD + TELERADIOLOGY FOR MODEL ENHANCEMENT

Edge device-based learning, coupled with cloud-based machine learning, is a unique model. A system's reliability becomes difficult to question when cloud services are used in conjunction with edge devices to eliminate break downs [16]. This model can be further strengthened by a concept called "federated learning" having the best features from both systems. Coined by Google researchers, federated learning refers to machine learning models learning on private data without moving the data from its source and compromising privacy [17]. Edge devices can employ this to give an instant diagnosis through narrow AI and run processes that require low computation power. Cloud-based tools can use their high capacity computing power to run additional algorithms on the same X-ray image, add to a list of differential diagnoses, detect other disease or comorbidity conditions, and run image analytics.

The addition of a human-in-the-loop through the teleradiology model can help in close monitoring of the performance of the edge and cloud devices and ensure that no drifting takes place. Teleradiology involves transferring radiological data from one location to the other by the use of some telecommunication systems [18]. In the hybrid model, a medical imaging specialist keeps track of both the models and validates the results. The imaging expert in the loop can edit and update reports and recategorize studies, giving valuable feedback to the system by an embedded AI feedback loop. The cloud teleradiology model has a provision for bringing outlier cases back in the workflow, as imaging experts can recheck the results, thus minimizing the chances of the algorithm overlooking noisy images harboring a subtle disease. The feedback given by imaging experts can help train the models and thereby completing a critical feedback loop.

We know that narrow AI functions can be performed effortlessly using the currently available CPUs, but running multiple algorithms on edge devices can be challenging. Developments in hardware design are changing this landscape, as AI chipsets with multiple cores which are ideal for parallel computing, like the NVIDIA A100 GPU, can handle the new design and computation challenges [19].

## X. FUTURE

Edge diagnostics and edge learning devices are going to power the devices of the future in the medical imaging space. In their current form, these devices are equipped to perform basic analysis such as triage and pre-screen studies such as normal-abnormal, pneumonia-healthy, and TB-no TB on X-rays, and bleed-no bleed on brain CT scans. As high computing power makes deep learning solution deployment easier on edge devices, the future of edge computing is likely to get refined further as AI chipsets undergo major shifts in design. Multi-class models are likely be deployed on edge devices with built-in localization and quantification models. The bouquet of functions on edge devices is likely to grow multifold. The democratization of radiology could also be a reality if edge devices are deployed to perform basic triage functions in remote areas where medical imaging specialists are unavailable. Delivering radiology and diagnostics with a negligible wait time for triage results would make these truly point-of-care. After initial triage, human expert readers can look at these studies and decide the next course of action. Edge learning is a unique opportunity to support basic health services and public and population health since this technology can be enabled as easily in larger setups such as primary, secondary, tertiary, and quaternary care hospitals as it can be deployed in MDUs. It is to be noted that the use case in these settings will differ. For example, in a hospital setting, this technology can decongest the workflow of imaging experts and help them focus on important studies and emergencies first, before moving on to normal studies. More solutions and advanced workflows are likely to emerge out of these scenarios which will define imaging protocols in the near future.

## CONFLICT OF INTEREST STATEMENT

Dr Amit Kharat is a Professor of Radiology at DPU, Pune. Dr Vinay Duddalwar is a consultant for Radmetrix, Intuit Systems and a member of the advisory board for DeepTek. He has been awarded a research grant by Samsung Healthcare and is in a research collaboration with Fujifiilm Medical Systems, Philips Healthcare.

AUTHORS

**First Author** – Dr. Amit Kharat, MBBS, DMRD, DNB, PhD, FICR, DeepTek Inc, amit@deeptek.ai
**Second Author** – Dr Vinay Duddalwar, MD, FRCR, University of Southern California, Vinay.Duddalwar@med.usc.edu
**Third Author** – Dr Krishna Saoji, MD, Dr D Y Patil Medical College, Pune, KRISHNASAOJI@gmail.com

**Correspondence Author** – Dr. Amit Kharat, MBBS, DMRD, DNB, PhD, FICR, DeepTek Inc, amit@deeptek.ai.